\documentclass[10pt, conference, compsocconf,letterpaper,twocolumn]{IEEEtran}

\usepackage[rgb]{xcolor}
\usepackage[subrefformat=parens,labelformat=parens]{subfig}
\usepackage{caption,url}
\usepackage{moreverb, amsmath, amsfonts}
\usepackage{graphicx}
\usepackage{placeins}
\usepackage[ruled]{algorithm2e}
\usepackage{algorithmic}
\usepackage{booktabs}
\usepackage{multirow}
\usepackage{xspace}
\usepackage{todonotes}
\usepackage{adjustbox}
\usepackage{scalefnt}

\usepackage{pgf}
\newcommand{\ignore}[1]{}

\newcommand{\CS}{\ensuremath{\mathcal C}\xspace}
\newcommand{\JS}{\ensuremath{\mathcal J}\xspace}
\newcommand{\IS}{\ensuremath{\mathcal I}\xspace}
\newcommand{\KS}{\ensuremath{\mathcal K}\xspace}

\newcommand{\LS}{\ensuremath{\mathcal L}\xspace}
\newcommand{\MS}{\ensuremath{\mathcal M}\xspace}
\newcommand{\NS}{\ensuremath{\mathcal N}\xspace}

\newcommand{\etc}{\textit{etc}.\xspace}

\newcommand{\dgdft}{\textsf{DGDFT}\xspace}
\newcommand{\siesta}{\textsf{SIESTA}\xspace}

\newcommand{\superlu}{\textsf{SuperLU\_DIST}\xspace}

\title{A Left-Looking Selected Inversion Algorithm and Task Parallelism on Shared Memory Systems} 

\author{
\IEEEauthorblockN{Mathias Jacquelin\IEEEauthorrefmark{1}, Lin Lin\IEEEauthorrefmark{2}\IEEEauthorrefmark{1}, Weile Jia\IEEEauthorrefmark{3}, Yonghua Zhao\IEEEauthorrefmark{3}, Chao Yang\IEEEauthorrefmark{1}}
\IEEEauthorblockA{\IEEEauthorrefmark{1}
Lawrence Berkeley National Laboratory\\
\url{mjacquelin@lbl.gov}, \url{cyang@lbl.gov}}
\IEEEauthorblockA{\IEEEauthorrefmark{2}
University of California Berkeley\\
\url{linlin@math.berkeley.edu}}
\IEEEauthorblockA{\IEEEauthorrefmark{3}
Supercomputing Center of Chinese Academy of Sciences\\
Beijing, China\\
\url{jiawl@sccas.cn}, \url{yhzhao@sccas.cn}}
}

\begin{document}

\maketitle

\begin{abstract}
Given a sparse matrix $A$, the selected inversion algorithm is an efficient method for computing certain selected elements of  $A^{-1}$. These selected elements correspond to all or some nonzero elements of the $LU$ factors of $A$. 
In many ways, the type of matrix updates performed in the selected 
inversion algorithm is similar to that performed in the $LU$ factorization,
although the sequence of operation is different.
In the context of $LU$ factorization, it is known that the left-looking and right-looking algorithms exhibit different memory access and data communication patterns, and hence different behavior on shared memory and distributed memory parallel machines. %In particular, the right-looking $LU$ factorization can reach higher parallel scalability on large-scale distributed memory parallel machines. 
Corresponding to right-looking and left-looking LU factorization, selected inversion algorithm can be organized as a left-looking and a right-looking algorithm. 
%From the point of view of the elimination tree, the $LU$ factorization and 
%the selected inversion traverse upward and downward along the elimination tree,
%respectively. Hence the selected inversion algorithm can be viewed as the $LU$ f%actorization performed in the ``mirrored'' direction. 
The parallel right-looking version of the algorithm has been developed in~\cite{JacquelinLinYang2015}. 
The sequence of operations performed in this version of the selected
inversion algorithm is similar to those performed in a left-looking $LU$ 
factorization algorithm. 
In this paper, we describe the left-looking variant of the selected 
inversion algorithm, and based on task parallel method, present an efficient implementation of the algorithm for shared memory machines. We demonstrate that with the task scheduling features provided by OpenMP 4.0, the left-looking selected inversion algorithm can scale well both on the  Intel Haswell multicore architecture and on the Intel Knights Corner (KNC) manycore architecture. Compared to the right-looking selected inversion algorithm, the left-looking formulation  facilitates pipelining of work along different branches of the elimination tree, and can be a promising candidate for future development of massively parallel selected inversion algorithms on heterogeneous architecture. 
\end{abstract}

%toms related stuff
%\category{G.4}{Mathematical Software}{}[Algorithm design and analysis]
% * <linlin.pku@gmail.com> 2016-01-20T17:53:18.716Z:
%
% ^.
%\category{G.4}{Mathematical Software}{}[Parallel and vector implementations]
%\category{I.1.2}{Symbolic and Algebraic Manipulation}{Algorithms}[Algebraic algorithms]
%\terms{Design, Performance}
%\keywords{}


\section{Introduction}

%\LL{Mathias: how to put affiliation in this format? Affiliation for Weile and Yonghua:Supercomputing Center of Chinese Academy of Sciences, Beijing, China}

Given a non-singular, sparse matrix $A\in \mathbb{C}^{N\times N}$, the selected inversion algorithm is an efficient method for computing selected elements of $A^{-1}$. These selected elements correspond to all or a subset of the nonzero entries of the $LU$ factors of $A$. The computation of such selected elements has recently received an increasing level of attention, notably in the context of density functional theory~\cite{HohenbergKohn1964,KohnSham1965,LinLuYingCarE2009,LinChenYangEtAl2013}, quantum transport~\cite{LiAhmedKlimeckDarve2008,LiWuDarve2013}, dynamical mean field theory~\cite{KotliarSavrasovHauleEtAl2006,TangSaad2012}, and uncertainty quantification~\cite{BekasCurioniFedulova2009}, to name a few. 

The sequence of operations performed in the selected inversion of $A$ can be 
described in terms of a traversal of the elimination tree associated with
$A$. Elimination tree traversal can also be used to describe the sequence of 
operations performed in an $LU$ factorization of $A$. However, in selected 
inversion, the elimination tree is traversed from the root down to the leaves, 
whereas a bottom-up traversal from the leaves to the root is performed
in the $LU$ factorization. Hence, the sequence of operations performed
in the selected inversion of $A$ can be viewed as ``mirrored" operations 
performed in the $LU$ factorization of $A$. 

%The reason why selected inversion algorithm works can be most clearly
%viewed from the elimination tree~\cite{Liu1990} corresponding to the $LU$ factorization. The $LU$ factorization starts from the bottom (a.k.a. ``leaf'') of the elimination tree, and traverses upward till reaching the top (a.k.a ``root'') of the tree. Correspondingly, the selected inversion algorithm starts from the root, and traverses downward the tree till reaching all leaves. The efficiency of the selected inversion algorithm lies in the fact that all computation involves only elements ``on the tree'', i.e. elements related to the sparsity pattern of the $LU$ factors.  Hence the selected inversion algorithm can be viewed as the $LU$ factorization performed in the ``mirrored'' direction.

There are several ways to implement the $LU$ or Cholesky factorization of $A$. 
Two of the most widely used implementations are the left-looking and 
right-looking factorization algorithms. 
They differ in the way data is fetched from the factored part of matrix 
and applied to the part of the matrix that remains to be factored.

%Their differences can be summarized as follows. Once the computation of a column or a supernode is finished, 

%left-looking algorithms tend to maximally postpone the update of the trailing matrix
%, until necessary when factorizing a column in the trailing matrix.
%Hence when factorizing any column, the algorithm always fetches information from the computed parts of the matrix, i.e. ``from the left''. On the other hand, right-looking algorithms tend to maximally update the trailing matrix, and move on to a new column until no further update can be performed. Hence the algorithm always pushes information to the trailing matrix, i.e. ``to the right''.  

It is well known that the left-looking and right-looking algorithms 
exhibit different  memory access and data communication patterns. 
As a result, their performance can be quite different on shared memory 
and distributed memory parallel machines. The right-looking $LU$ 
factorization can sometimes achieve higher parallel scalability 
on distributed memory parallel machines with a relatively large number 
of processors~\cite{LiDemmel2003,YamazakiLi2012}.  

There are at least two ways to implement the selected inversion algorithm. The right-looking variant has been developed in~\cite{LinYangMezaEtAl2011}. 
Its parallelization for a distributed memory machine has been  
described in~\cite{JacquelinLinYang2015}.  Although the parallel right-looking 
algorithm can scale to as many as 4,096 processors~\cite{JacquelinLinWichmannEtAl2015},
further performance improvement appears to be challenging
due to the complex data communication patterns employed in this
variant of the selected inversion algorithm.

% which always fetch information from the computed parts of $A^{-1}$, which appear to the right of the current supernode.  
%In this sense, such algorithms can be viewed as ``right-looking'' algorithms. We have recently demonstrated that the right-looking selected inversion algorithm can exhibit high parallel scalability on massively parallel distributed memory machines with over $4096$ cores~\cite{JacquelinLinYang2015}. 

%On the other hand, further improvement of the parallel scalability of right-looking selected inversion algorithm can be  limited by the relatively complex data communication pattern: all computed parts of $A^{-1}$ can be fetched repeatedly, even when the current supernode is a leaf node. This significantly complicates the scheduling procedure for exploiting parallelism~\cite{JacquelinLinYang2015}.  

%There is an additional style of selected inversion based on the multifrontal method~\cite{AmestoyDuffLExcellentEtAl2012a,LinYangLuEtAl2011}. The algorithmic structure and implementation of multifrontal algorithms are significantly different from the left-looking and right-looking algorithms. Multifrontal algorithms can achieve high parallel scalability for matrices with balanced elimination tree structure, but the load balancing can be challenging for matrices with unbalanced tree structure, which limits the parallel scalability.

In this paper, we present a left-looking selected inversion algorithm. 
We will show that it is much easier to schedule multiple tasks 
that can be executed concurrently in the left-looking algorithm.
As a result, the left-looking implementation may reach higher 
parallel scalability than what is possible at present.

%Similar to the right-looking $LU$ factorization, this algorithm maximally pushes information towards the leaves of the tree to be computed in the future.  
%The advantage of the left-looking algorithm is that once a column or a supernode of $A^{-1}$ finishes its computation, its information will never be fetched to again in future computation, and therefore the scheduling algorithm can be significantly simplified. This brings the possibility of further extending the parallel scalability of selected inversion algorithm for large scale matrices in the massively parallel computing environment.

%On the other hand, as supercomputer nodes grow ``fatter''
%with multicore and manycore processors in the High Performance Computing world,	the future of software development for supercomputing relies increasingly on high
%level programming models such as OpenMP for on-node parallelism.
As a first step, we develop an efficient implementation of the left-looking 
selected inversion algorithm for shared memory parallel machines. 
The parallelization makes use of the task scheduling features provided by 
OpenMP 4.0. We demonstrate the performance of our implementation on a number 
of test problems. The performance study is carried out on both the Intel 
Haswell multicore architecture and the Intel Knights Corner (KNC) manycore 
architecture. 

The rest of the paper is organized as follows. In the next section, we review the basic algorithmic ingredients of a selected inversion algorithm, and point out the main differences in different variants of the algorithm. We also describe the left-looking selected inversion in detail. In section~\ref{sec:parallel}, we discuss how various updates performed in a left-looking selected inversion can be divided as individual tasks,  how these tasks depend on each other, and how we can use dependency analysis to avoid write conflict. We also show how the execution of different tasks can be scheduled
dynamically based on dependency analysis,and how task scheduling can be implemented with the new OpenMP primitives. The numerical results that demonstrate the efficiency of the left-looking algorithm are presented in section~\ref{sec:example}.
%For instance, for relatively dense matrices such as DGDFT\_Graphene720 and nd3k, the left-looking selected inversion algorithm yields 

% Let $A\in \mathbb{C}^{N\times N}$ be a non-singular sparse matrix.  
% We use $A_{i,j}$ to denote the $(i,j)$-th entry of the matrix $A$, 
% and $A_{i,*}$ and $A_{*,j}$ to denote the $i$-th row and the $j$-th column
% of $A$, respectively.  

% We are interested in computing 
% {\em selected elements} of $A^{-1}$, defined as
% \begin{equation}
%   \{(A^{-1})_{i,j}\vert \ \ \mbox{for} \ \ 1\le i,j\le N, \ \ 
%   \mbox{such that} \ \ A_{i,j}\ne 0 \}.
%   \label{eqn:selelem}
% \end{equation}
% Sometimes, we only need to compute a subset of these selected elements,
% for example, the diagonal elements of $A^{-1}$. Computing the full
% inverse of $A$ and then extract the selected elements can be 
% often prohibitively expensive in practice. 
% If a sparse $LU$ factorization  of $A$ is available
% (or $LDL^{T}$ factorization if $A$ is symmetric) , a more efficient way 
% to achieve this goal is to use an algorithm that makes efficient use 
% of the sparse $L$ and $U$ factors of $A$.  In such an algorithm, 
% which we call selected inversion ({\em SelInv}), some 
% additional elements of $A^{-1}$ may need to be computed. However, the overall 
% set of nonzero elements that need to be computed often remains a small 
% percentage of all elements of $A^{-1}$ due to the spars ity structure of $A$.

\section{Theory}

\subsection{Selected inversion algorithm}

The selected inversion algorithm has been discussed
                                    in~\cite{LinYangMezaEtAl2011,JacquelinLinYang2015}, and here
we only briefly recall its formulation.
To simplify the discussion as well as its implementation, in this paper, we
assume the matrix $A$ is at least \textit{structurally symmetric}, i.e.
$A_{ij}\ne 0$ implies $A_{ji}\ne 0$ for any $i,j$. If $A$ is not
structurally symmetric we can fill zeros to the matrix $A$ and treat
these added zeros as nonzero entries, so that the resulting modified
matrix becomes structurally symmetric.  Given a 2-by-2 block
partitioning of the matrix $A$  with $A_{1,1}$ being a scalar, 
\begin{equation}
	A = \begin{pmatrix}
		A_{1,1} & A_{1,2}\\
		A_{2,1} & A_{2,2}
	\end{pmatrix},
	\label{}
\end{equation}
its $LU$ decomposition is 
\begin{equation}
	A = \begin{pmatrix}
		L_{1,1} & 0\\
		L_{2,1} & I 
	\end{pmatrix}
	\begin{pmatrix}
		U_{1,1} & U_{1,2}\\
		0      & S_{2,2}
	\end{pmatrix}.
	\label{eqn:LU2by2}
\end{equation}
Here $S_{2,2}$ is called the Schur complement, and is obtained from the
trailing submatrix of column 1, denoted by $A_{2,2}$, modified by a rank one
matrix from the $L,U$ factors.
We can express $A^{-1}$ as
\begin{equation}
	A^{-1} = \begin{pmatrix}
    (\hat{U}_{1,1})^{-1} (\hat{L}_{1,1})^{-1} + \hat{U}_{1,2}
    S_{2,2}^{-1}
		\hat{L}_{2,1} & - \hat{U}_{1,2} S_{2,2}^{-1} \\
		-S_{2,2}^{-1} \hat{L}_{2,1}  & S_{2,2}^{-1}
	\end{pmatrix},
	\label{eqn:Ainv2by2normal}
\end{equation}
where
\begin{equation}
  \begin{array}{ll}
	\hat{L}_{1,1} = L_{1,1}, & \hat{U}_{1,1}=U_{1,1},\\
	\hat{L}_{2,1} = L_{2,1}(L_{1,1})^{-1}, & \hat{U}_{1,2} = (U_{1,1})^{-1}
	U_{1,2}.
  \end{array}
	\label{}
\end{equation}

Assume the inverse of the Schur complement $S_{2,2}^{-1}$ has already been computed, and denote by $\CS$ the set of indices
\begin{equation}
	\{i|\left(L_{2,1}\right)_{i} \ne 0\}.
	\label{}
\end{equation}  
Due to the structural symmetry property of $A$,  the set
$\{j|\left(U_{1,2}\right)_{j} \ne 0\}$ is identical to $\CS$.
The basic idea of the selected inversion algorithm is that in order to
update $A^{-1}_{1,1}$, we only need
the entries 
\begin{equation}
\left\{\left( S_{2,2}^{-1} \right)_{i,j} | i\in \CS, j\in \CS\right\}.
	\label{eqn:selectentry2x2}
\end{equation} 
Applying this principle recursively, we obtain a pseudo-code for
demonstrating this column-based selected inversion algorithm for symmetric matrix,
which is given in~\cite{LinYangMezaEtAl2011}.

In practice, a column-based sparse factorization and selected inversion
algorithm may not be efficient due to the lack of level 3 BLAS
operations.  For a sparse matrix $A$, the columns of $A$ and the $L$
factor can be partitioned into supernodes. A supernode is a maximal set
of contiguous columns $\JS=\{j,j+1,\ldots,j+s\}$ of the $L$ factor that
have the same nonzero structure below the $(j+s)$-th row, and the lower
triangular part of $L_{\JS,\JS}$ is dense. This definition can be
relaxed to limit the maximal number of columns in a supernode (i.e. sets
are not necessarily maximal).  With slight abuse of notation, both a
supernode index and the set of column indices associated with a
supernode are denoted by uppercase script letters such as $\IS,\JS,\KS$
\etc.  $A_{\IS,*}$ and $A_{*,\JS}$ are used to denote the $\IS$-th block
row and the $\JS$-th block column of $A$, respectively.
$A_{\IS,\JS}^{-1}$ denotes the $(\IS,\JS)$-th block of the matrix
$A^{-1}$, i.e. $A_{\IS,\JS}^{-1}\equiv (A^{-1})_{\IS,\JS}$.  When the
block $A_{\IS,\JS}$ itself is invertible, its inverse is denoted by
$(A_{\IS,\JS})^{-1}$ to distinguish from $A_{\IS,\JS}^{-1}$.
Using the supernode notation, a pseudo-code for the selected inversion
algorithm is given in Algorithm~\ref{alg:selinvlu}.   Here $L_{\IS,\KS}\ne 0$ means that it is not an empty matrix block.

%\LL{Talk about supernode and elimination tree}

\begin{algorithm}[ht]
  \small
  \DontPrintSemicolon
  \caption{Selected inversion algorithm based on $LU$ factorization.}
  \label{alg:selinvlu}

  \KwIn{\begin{tabular}{l} (1) \begin{minipage}[t]{2.5in} The supernode partition of columns of $A$: $\{1,2,...,\mathcal{N}\}$ \end{minipage}\\
        (2) \begin{minipage}[t]{2.5in} A supernodal $LU$ factorization of $A$ with $LU$ factors $L$ and $U$.  \end{minipage}
        \end{tabular}
        }

            %(3) \begin{minipage}[t]{4.0in}
   \KwOut{\begin{minipage}[t]{2.5in} Selected elements of $A^{-1}$, i.e. $A^{-1}_{\IS,\JS}$ such
             that $L_{\IS,\JS}$ is not an empty block. \end{minipage}
        } 

	\For{$\KS = \mathcal{N}, \mathcal{N}-1, ..., 1$}{
    \lnl{alg1.step0} Find the collection of indices\;
    $\CS=\{\IS~|~\IS>\KS,L_{\IS,\KS}\ne 0\}$\;
    \lnl{alg1.step1} $\hat{L}_{\CS,\KS}\gets L_{\CS,\KS} (L_{\KS,\KS})^{-1},
    \hat{U}_{\KS,\CS}\gets (U_{\KS,\KS})^{-1} U_{\KS,\CS}$\; 
  }

	\For{$\KS = \mathcal{N}, \mathcal{N}-1, ..., 1$}{
    Find the collection of indices\;
    $\CS=\{\IS~|~\IS>\KS,L_{\IS,\KS}\ne 0\}$\;
	  \lnl{alg1.step2} Calculate $A^{-1}_{\CS,\KS} \gets -A^{-1}_{\CS,\CS}
	  \hat{L}_{\CS,\KS}$\; 
	  \lnl{alg1.step3} Calculate $A^{-1}_{\KS,\KS} \gets U_{\KS,\KS}^{-1}
  	L_{\KS,\KS}^{-1} - \hat{U}_{\KS,\CS} A^{-1}_{\CS,\KS}$\; 
	  \lnl{alg1.step4} Calculate $A^{-1}_{\KS,\CS} \gets - \hat{U}_{\KS,\CS}
	  A^{-1}_{\CS,\CS}$\;
  }
\end{algorithm}

\subsection{Left-looking, right-looking, and multifrontal algorithms}

There are three main variations of an $LU$ or $LDL^T$ 
factorization algorithms. They are the left-looking, right-looking 
and multifrontal
algorithms~\cite{EisenstatSchultzSherman1981,GeorgeHeathLiuEtAl1988,NgPeyton1993,LiDemmel2003,DuffReid1983}.
The difference among these approaches lies mainly in the way the
Schur complement is updated. In the left-looking
algorithm, the update of the $\JS$-th supernode within the Schur complement
is delayed until the supernodes $\KS$ of $L$ (and $U$) have been 
computed for all $\KS < \JS$.  When the $\JS$-th supernode is updated,
the updating procedure looks to the left of the $\JS$-th supernode, 
and collects contributing matrix blocks from supernodes $\KS$ with $\KS < \JS$. 
%Hence the name left-looking. 
The collected contributing matrix blocks are accumulated by means of matrix inner products. 

In the right-looking algorithm, the entire Schur complement to the 
right of $\JS$-th supernode is updated when the $\JS$-th supernode 
of $L$ becomes available.  The update is performed as a matrix 
outer product of the $L$ and $U$ factors from the $\JS$-th supernode. 

The multifrontal algorithm can be considered as an variant of the 
right-looking algorithm. In a multifrontal algorithm, the update
of the Schur complement is organized in a hierarchical fashion, 
and guided by the elimination tree~\cite{Liu1990} that describes
the dependency among all supernodes in the $LU$ factorization.  
The hierarchical update requires the contributions of a supernode 
to its ancestors to be kept on a stack.  
%is visited, the update matrices from %its children on the elimination tree are removed from the top of the
%stack and assembled.  Then new update matrices are placed at the top of
%the stack. 

The left-looking, right-looking and multifrontal algorithms all have
advantages and disadvantages over each other. Their relative 
performance depends on the sparsity structure of the matrix 
and the architecture of the machine on which they are performed.
We refer readers to references~\cite{RothbergGupta1993} on comparisons of these algorithms
in the context of $LU$ or $LDL^T$ factorizations.

From the perspective of the elimination tree, an 
$LU$ factorization traverses from the bottom (leaf nodes) of the tree upwards until reaching the top (root node).  
The selected inversion algorithm shown in 
Alg.~\ref{alg:selinvlu}, on the other hand, can be described in terms 
of a top-down traversal of the elimination tree.
In Alg.~\ref{alg:selinvlu}, the main computational
bottlenecks are steps 3 and 5. In order to compute $A^{-1}_{\CS,\KS}$,
contributing blocks need to be fetched from the trailing submatrix 
$A^{-1}_{\CS,\CS}$, which is to the right the supernode $\KS$. 
In this sense, the most straightforward implementation of 
the selected inversion algorithm is a right-looking algorithm.
The contributions to the update of $A^{-1}_{\CS,\JS}$, are 
accumulated as an inner product between row blocks of $A^{-1}_{\CS,\CS}$ 
and $\hat{L}_{\CS,\JS}$ (or $\hat{U}_{\JS,\CS}$). 
%This type of
%update is similar to the update of the $\JS$th supernode of $L$ 
%in the left-looking $LU$ factorization of $A$.  
%Therefore, Alg.~\ref{alg:selinvlu} can also be viewed as a ``mirrored''
%left-looking algorithm, with the supernode $\JS$ serving as the
%``reflection plane''.  \CY{notation issue, should we use $\JS$ 
%instead of $\KS$ to be consistent with factorization? Do we really
%need to use the term 'mirror'. Selected inversion is already 
%a clear distinction from factorization. Is right-looking
%a right-looking algorithm?}

The implementation details of a sequential right-looking 
selected inversion algorithm for general sparse symmetric
matrices have been described in ~\cite{LinYangMezaEtAl2011}.
In~\cite{JacquelinLinYang2015,JacquelinLinWichmannEtAl2015},
we presented a parallel implementation of this right-looking algorithm. Our numerical experiments indicate 
that such an algorithm can scale to $4096$ or more processors. 

The selected inversion algorithm can also be implemented in a way
that is analogous to the multifrontal method used for a $LU$ factorization
of $A$.  This variant of the selected inversion algorithm is described
in~\cite{LinLuYingCarE2009}, which was referred to as a hierarchical 
Schur complement method. The parallel implementation 
of such method for a Laplacian type of matrices was presented in
~\cite{LinYangLuEtAl2011}. However, a load-balanced
implementation of this approach on massively parallel computers for general sparse matrices can become challenging.

\subsection{Left-looking selected inversion algorithm}

We now describe the left-looking variant of the selected inversion algorithm. 
This variant offers some advantages in terms of
load-balancing, memory access patterns and scheduling compared to the
other variants on massively parallel computer architectures.

In the left-looking selected inversion algorithm, when the computation for the supernode $\KS$ is finished
and $A_{\CS,\KS}^{-1}$ 
becomes available,  
we update 
%all descendants of $\KS$, 
all matrix blocks of $A^{-1}$ corresponding to the  descendants of $\KS$ within the nonzero sparsity pattern of the $LU$ factors.
This type of update is motivated by 
the right-looking factorization algorithm in which all ancestors of
 $\KS$ corresponding to the nonzero sparsity pattern of the $LU$ factor are updated, when $L_{\CS,\KS}$ and $U_{\KS,\CS}$ become available. 
%\MJ{This is not entirely true, we only update descendants of $\KS$ that have nonzero elements in the rows facing supernode $\KS$. Same thing for factorization: only SOME ancestors are updated. 
%This is mentioned a few paragraphs later, in order to introduce eq (9) but I'm wondering if we can state that.}\LL{Is this clarified now?}

To be specific, let us consider the update of the lower triangular part 
of $A^{-1}$ first. Define the sets
\begin{equation}
  \CS=\{\IS~|~\IS> \KS,L_{\IS,\KS}\ne 0\}, \quad \CS'=\{\IS~|~\IS< \KS,L_{\KS,\IS}\ne 0\},
  \label{eqn:newset}
\end{equation}
and the computation for the supernode $\KS$ is finished when $A_{\CS,\KS},A_{\KS,\CS}$ and $A_{\KS,\KS}$ are computed.
When the matrix blocks $A^{-1}_{\CS,\KS}$ become available, according to step 3 of Alg.~\ref{alg:selinvlu}, we can apply
updates to matrix blocks indexed by $\CS$ and $\CS'$ as follows
\begin{equation}
  A^{-1}_{\CS,\CS'} \leftarrow  A^{-1}_{\CS,\CS'} - A^{-1}_{\CS,\KS}
  \hat{L}_{\KS,\CS'}.
  \label{eqn:outerprodtmp}
\end{equation}
The update described by Eq.~\eqref{eqn:outerprodtmp} is clearly a block outer 
product. This is similar to the outer product used to update the Schur
complement in a right-looking factorization algorithm.
However, we should note that not all matrix blocks of $A^{-1}_{\CS,\CS'}$ 
need to be updated in the selected inversion algorithm. Only the 
matrix blocks of $A^{-1}_{\IS,\IS'}$ such that $\IS\in\CS, \IS'\in\CS'$
and $L_{\IS,\IS'} \ne 0$ need to be updated. Therefore, to be precise,
Eq.~\eqref{eqn:outerprodtmp} should be replaced by
\begin{equation}
\begin{adjustbox}{scale=0.9}
$
A^{-1}_{\IS,\IS'} \leftarrow  A^{-1}_{\IS,\IS'} - A^{-1}_{\IS,\KS}
  \hat{L}_{\KS,\IS'}, \quad \IS\in\CS,\IS'\in\CS',L_{\IS,\IS'}\ne 0.
$
  \label{eqn:outerprod}
\end{adjustbox}
\end{equation}

We should note that all the blocks below and to the right of $A^{-1}_{\KS,\KS}$
should have been computed when the $(\KS+1)$-th supernode has been traversed.
Hence, to complete the computation for the
$\KS$-th supernode, only the diagonal block $A^{-1}_{\KS,\KS}$
needs to be updated.  This is the first update performed in the
second loop of Alg.~\ref{alg:selinvlu}.

Once $A^{-1}_{\KS,\KS}$ becomes available, we can also update the
$\KS$-th block row of $A^{-1}$ by
\begin{equation}
\begin{adjustbox}{scale=0.9}
$
  A^{-1}_{\KS,\IS'} \leftarrow A^{-1}_{\KS,\IS'} - \sum_{\substack{\IS\in\CS,\\
  L_{\IS,\IS'\ne 0}}} A^{-1}_{\KS,\IS}
  \hat{L}_{\IS,\IS'}, \quad \IS'\in \CS'.
$
  \label{eqn:innerprod}
\end{adjustbox}
\end{equation}
The update performed in Eq.~\eqref{eqn:innerprod} is a block inner product
calculation.

%When $A^{-1}_{\CS,\KS}$ becomes available, $A^{-1}_{\KS,\CS}$ is also readily available to update supernodes to the left of the supernode $\KS$. Unlike, $A^{-1}_{\CS,\KS}$, the matrix blocks $A^{-1}_{\KS,\CS}$ only updates $A^{-1}_{\KS,\CS'}$ in the lower triangular part of $A^{-1}$.
%Again the sparsity pattern of $L$ should be properly included as
%\begin{equation}
%  A^{-1}_{\KS,\IS'} \leftarrow A^{-1}_{\KS,\IS'} - \sum_{\substack{\IS\in\CS,\\
%  L_{\IS,\IS'\ne 0}}} A^{-1}_{\KS,\IS}
%  \hat{L}_{\IS,\IS'}, \quad \IS'\in \CS'.
%  \label{eqn:innerprod}
%\end{equation}
%Eq.~\eqref{eqn:innerprod} resembles an inner product between
%$A^{-1}_{\KS,\CS}$ and $\hat{L}_{\CS,\IS'}$, but the condition
%$L_{\IS,\IS'}$ should be properly included to reduce the computational
%cost.
%
%Finally, the diagonal blocks $A^{-1}_{\KS,\KS}$ also updates
%$A^{-1}_{\KS,\CS'}$ as
%\begin{equation}
%  A^{-1}_{\KS,\IS'} \leftarrow A^{-1}_{\KS,\IS'} -  A^{-1}_{\KS,\KS}
%  \hat{L}_{\KS,\IS'}, \quad \IS'\in \CS'.
%  \label{eqn:diag}
%\end{equation}
%Eq.~\eqref{eqn:diag} has a similar structure as~\eqref{eqn:innerprod},
%and can be combined together into the inner product phase~\eqref{eqn:innerprod} in practical
%implementation. 

The update to the upper triangular blocks of $A^{-1}$ can be performed
in a similar fashion.
The pseudo-code that outlines the main steps of the sequential 
left-looking selected inversion algorithm is given 
in Alg.~\ref{alg:selinvluleft}.

\begin{algorithm}[h]
  \small
  \DontPrintSemicolon
  \caption{Left-looking selected inversion algorithm based on $LU$ factorization.}
  \label{alg:selinvluleft}

  \KwIn{\begin{tabular}{l} (1) \begin{minipage}[t]{2.5in} The supernode partition of columns of $A$: $\{1,2,...,\mathcal{N}\}$ \end{minipage}\\
        (2) \begin{minipage}[t]{2.5in} A supernodal $LU$ factorization of $A$ with $LU$ factors $L$ and $U$.  \end{minipage}
        \end{tabular}
        }

            %(3) \begin{minipage}[t]{4.0in}
   \KwOut{\begin{minipage}[t]{2.5in} Selected elements of $A^{-1}$, i.e. $A^{-1}_{\IS,\JS}$ such
             that $L_{\IS,\JS}$ is not an empty block. \end{minipage}
        } 

	\For{$\KS = \mathcal{N}, \mathcal{N}-1, ..., 1$}{
    \lnl{alg1.step0}Find the collection of indices\;
    $\CS=\{\IS~|~\IS>\KS,L_{\IS,\KS}\ne 0\}$\;
    \lnl{alg1.step1}$\hat{L}_{\CS,\KS}\gets L_{\CS,\KS} (L_{\KS,\KS})^{-1},
    \hat{U}_{\KS,\CS}\gets (U_{\KS,\KS})^{-1} U_{\KS,\CS}$\; 
  }

  Set $A^{-1}$ to be a zero sparse matrix, with sparsity pattern given
  by $L+U$\;
	\For{$\KS = \mathcal{N}, \mathcal{N}-1, ..., 1$}{
    Find the collection of indices\;
    $\CS=\{\IS~|~\IS>\KS,L_{\IS,\KS}\ne 0\}$\;
	  \lnl{alg1.step2}Update the matrix diagonal block $A^{-1}_{\KS,\KS} \gets U_{\KS,\KS}^{-1}
  	L_{\KS,\KS}^{-1} - \hat{U}_{\KS,\CS} A^{-1}_{\CS,\KS}$\; 
    Find the collection of indices\;
    $\CS'=\{\IS~|~\IS< \KS,L_{\KS,\IS}\ne 0\}$\;
    \lnl{alg1.step3}Outer product phase for the lower triangular part:\;
    \begin{multline}
    %\[
    A^{-1}_{\IS,\IS'} \leftarrow  A^{-1}_{\IS,\IS'} - A^{-1}_{\IS,\KS}
    \hat{L}_{\KS,\IS'},\\ \mbox{for}\quad
    \IS\in\CS,\IS'\in\CS',L_{\IS,\IS'}\ne 0.
    %\]
    \end{multline}
    \lnl{alg1.step4}Inner product phase for the lower triangular part:\;
    \[
    A^{-1}_{\KS,\IS'} \leftarrow A^{-1}_{\KS,\IS'} -
    \sum_{\substack{\IS\in\KS\cup \CS,\\
    L_{\IS,\IS'\ne 0}}} A^{-1}_{\KS,\IS}
    \hat{L}_{\IS,\IS'}, \quad \mbox{for}\quad \IS'\in \CS'
    \]
    \lnl{alg1.step3}Outer product phase for the upper triangular part:\;
    \begin{multline}
    %\[
    A^{-1}_{\IS',\IS} \leftarrow  A^{-1}_{\IS',\IS} -
    \hat{U}_{\IS',\KS}A^{-1}_{\KS,\IS},\\ \mbox{for}\quad
    \IS\in\CS,\IS'\in\CS',U_{\IS',\IS}\ne 0.
    %\]
    \end{multline}
    \lnl{alg1.step4}Inner product phase for the upper triangular part:\;
    \[
    A^{-1}_{\IS',\KS} \leftarrow A^{-1}_{\IS',\KS} -
    \sum_{\substack{\IS\in\KS\cup \CS,\\
    U_{\IS',\IS\ne 0}}} \hat{U}_{\IS',\IS} A^{-1}_{\IS,\KS}, \quad
    \mbox{for}\quad \IS'\in \CS'
    \]
  }
%\end{adjustbox}
\end{algorithm}

For symmetric matrices, the $LU$ factorization can be simplified into
the $LDL^{T}$ factorization. The update of upper triangular part in step 6 and 7 in Alg.~\ref{alg:selinvluleft} can simply
be obtained by the transpose of the lower triangular part without
further computation. As discussed in~\cite{JacquelinLinYang2015}, for
symmetric matrices, it is important to symmetrize the diagonal block
once $A^{-1}_{\KS,\KS}$ is computed in step 3 of
Alg.~\ref{alg:selinvluleft} as
\[
A^{-1}_{\KS,\KS} \gets \frac12 (A^{-1}_{\KS,\KS}+A^{-T}_{\KS,\KS}),
\]
in order to reduce the propagation of the symmetrization error. This is
particularly important for large matrices. The modification for
Hermitian matrices is similar, simply by replacing the transpose
operation into the Hermitian transpose operation whenever suitable.

At first sight, 
%by comparing  Alg.~\ref{alg:selinvluleft} and~\ref{alg:selinvlu}, 
 the left-looking selected inversion algorithm has some disadvantages compared to the right-looking variant. 
The order of operations of the two algorithms are very different, and the implementation of the left-looking algorithm is  more complicated.  Furthermore, the left-looking selected inversion algorithm could result in higher memory consumption. In the right-looking selected inversion algorithm, one can gradually overwrite the $LU$ factors by $A^{-1}$, and hence the $LU$ factor and the $A^{-1}$ can share the same memory space. On the other hand, each update of $A^{-1}$ in the left-looking algorithm requires both the $LU$ and the $A^{-1}$ matrix blocks. Hence the storage cost of the left-looking algorithm can be close to twice as large as that in the right-looking algorithm.

On the other hand, the left-looking algorithm can become advantageous in the massively parallel computational environment by exploiting concurrency more naturally. In order to facilitate parallelism in the right-looking selected inversion algorithm, a task scheduling procedure guided by the traversal of the elimination tree is used to pipeline multiple tasks~\cite{JacquelinLinYang2015}. However, it is difficult to optimize this task scheduling procedure in the right-looking algorithm. This is because when the computation of a given supernode $\KS$ is finished, the matrix blocks $A_{\CS,\KS}$ can be requested repeatedly by supernodes to the left of $\KS$ in later computational stages (see step 3 in Alg.~\ref{alg:selinvlu}). 
This creates complex task dependencies, and hinders parallelism on distributed parallel computer architecture. The left-looking algorithm, on the other hand, has the advantage that once the contributions from 
$A^{-1}_{\CS,\KS}$ and $A^{-1}_{\KS,\CS}$ have been included in the matrix blocks associated with the descendants of $\KS$,
$A^{-1}_{\CS,\KS}$ and $A^{-1}_{\KS,\CS}$ will no longer be needed in any  subsequent calculations. This can greatly simplify the task dependency, and allows the selected inversion algorithm to become more load balanced and scalable on massively parallel computers.

% is that it allows us to express 
% the computation performed at supernode $\KS$ in terms of concurrent tasks more easily. This high level of concurrency in this variant of the selected 
% inversion algorithm allows us to leverage hardware concurrency. 

% In~\cite{JacquelinLinYang2015}, a two-dimensional block-cyclic data distribution is employed for the right-looking algorithm. In order to facilitate parallelism, a task scheduling procedure guided by the traversal of the elimination tree is used to pipeline multiple tasks. However, the task scheduling procedure in the right-looking algorithm is intrinsically complicated, since the contribution from any completed supernode $\KS$ can be fetched indefinitely by the supernodes to the left of $\KS$ in step 3 and 5 in Alg.~\ref{alg:selinvlu}. \CY{not sure what this means} This creates complex task dependencies, especially on distributed parallel computer architecture. The left-looking algorithm, on the other hand, has the advantage that once the contributions from 
% $A^{-1}_{\CS,\KS}$ and $A^{-1}_{\KS,\CS}$ have been included in the matrix 
% blocks associated with the descendants of $\KS$,
% $A^{-1}_{\CS,\KS}$ and $A^{-1}_{\KS,\CS}$ will no longer be needed in any 
% subsequent calculations. This greatly simplifies the task dependency and allows the selected inversion algorithm to become more load balanced and scalable on massively parallel computers.

\section{Task based parallelism and OpenMP implementation}
\label{sec:parallel}

As supercomputer nodes grow ``fatter'' with multicore and manycore 
processors, the performance of an application relies increasingly 
on using high level programming models such as OpenMP to achieve 
intra-node parallelism. Due to the relatively complex data dependency in the 
selected inversion algorithm, simple parallelization strategies such as those 
based on multi-threaded BLAS or parallel \textsf{for} loops cannot achieve
satisfactory scalablity on manycore shared memory nodes. 
Scalable implementation of the selected inversion algorithm requires a careful organization of the computation into relatively independent computational tasks with properly described task dependency and granularity. 
In the following, we demonstrate how the left-looking algorithm 
can be parallelized on a shared memory node by using OpenMP to manage concurrent threads for symmetric matrices. The computational tasks and dependencies can be described relatively easily thanks to the \textsf{task} and \textsf{task dependency} feature in OpenMP 4.0. A distributed memory parallel implementation and a hybrid MPI+OpenMP version will be described in a separate report.

% the computational 
% For relatively simple 
% The recently added OpenMP constructs for task parallelism raise the level of abstraction to improve programmer productivity. This will be a welcome development for scientific computing as supercomputer nodes grow "fatter" with multicore and manycore processors. However, if the algorithm is not suited for task parallel efficiently on the available multicore or manycore systems, the benefits will be lost. An efficient task algorithm must need to exploit cache and memory locality, maintain load balance, and minimize overhead cost. 

%\LL{The labels such as $L,M,N$ as well as $O_I^K$ are not referred to / defined in the text.}

\subsection{Task based scheduling procedure}

% \LL{We might need to talk about the scheduling somewhere here in the context of MPI / OpenMP hybrid implementation. This is the advantage of the left-looking algorithm, but so far it only appears in the introduction.}

The left-looking selected inversion in Alg.~\ref{alg:selinvluleft} can be organized into a ``pre-selected inversion'' phase (step 1-2) and the ``selected inversion'' phase (step 3-7). The pre-selected inversion phase computes the normalized $LU$ factors $\hat{L}$ and $\hat{U}$, respectively. This can be performed independently for each supernode $\KS$ and its parallelization can be simply performed by means of a parallel \textsf{for} loop.

\begin{figure}[htbp]
\centering
\begin{adjustbox}{width=\linewidth}
\includegraphics{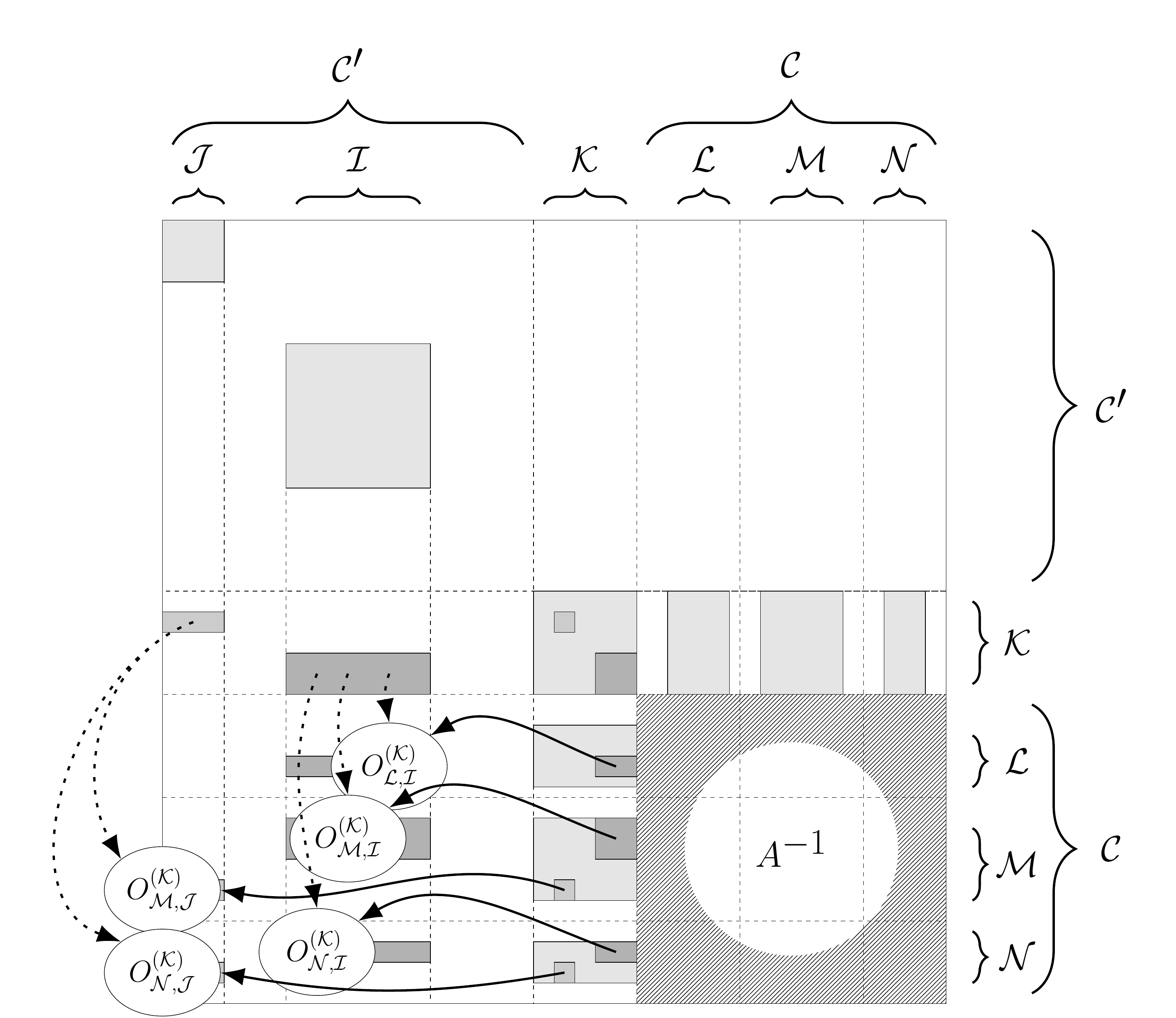}
\end{adjustbox}
\caption{Outer-product task parallelism. $O^{(\KS)}_{*,\IS}$ and $O^{(\KS)}_{*,\JS}$ correspond respectively to the outer-product updates from supernode \KS to two supernodes $\IS$ and $\JS$ in $\CS'$. Data dependencies from previously computed values of $A^{-1}$ are denoted with solid arrows. Data dependencies from values in $LU$ factors are indicated using dashed arrows.}
\label{fig:outer}
\end{figure}

\begin{figure}[htbp]
\centering
\begin{adjustbox}{width=\linewidth}
\includegraphics{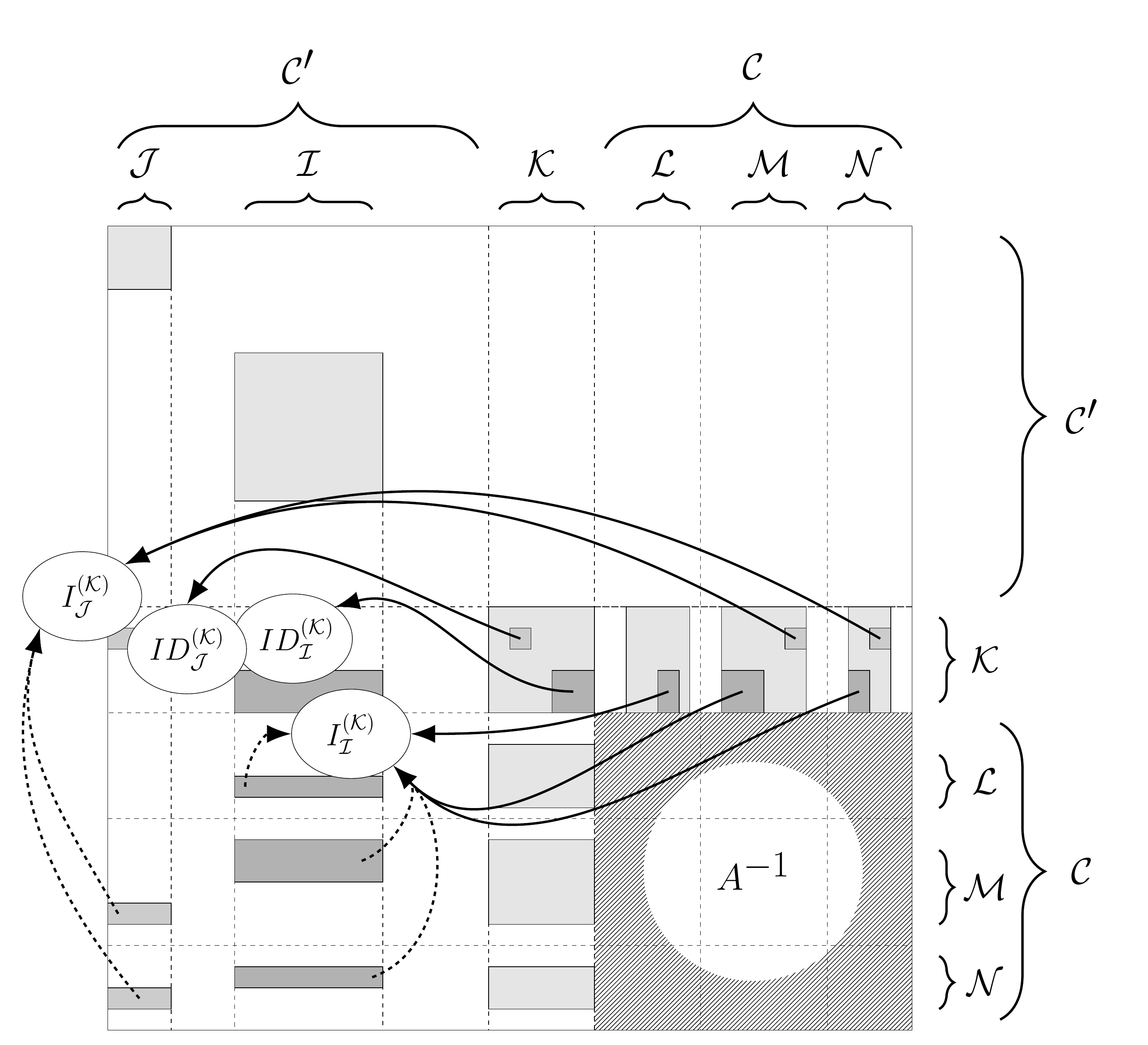}
\end{adjustbox}
\caption{Inner-product and update from diagonal task parallelism. $I^{(\KS)}_{\IS}$ and $I^{(\KS)}_{\JS}$ correspond to two inner-product updates from supernode \KS to two supernodes \IS and \JS in $\CS'$. $ID^{(\KS)}_{\IS}$ and $ID^{(\KS)}_{\JS}$ correspond to the updates from the diagonal block of \KS to those two supernodes in $\CS'$. Data dependencies from previously computed values of $A^{-1}$ are denoted with solid arrows. Data dependencies from values in $LU$ factors are indicated using dashed arrows.}
\label{fig:inner}
\end{figure}

The main difficulty is in the selected inversion phase. For each supernode $\KS$, the computation can be divided into 
three stages: (1) Diagonal block update (2) Outer-product update, and (3) Inner-product update.
In the following, and as depicted in Figures~\ref{fig:outer} and~\ref{fig:inner}, we assume that $\KS$ is the current
supernode being processed. Supernodes $\LS$, $\MS$ and
$\NS \in \CS$ are three supernodes that have already been 
computed, and they are ancestors of $\KS$ in the elimination tree. 
Supernodes $\JS$ and $\IS \in \CS'$ are
descendants of $\KS$ in the elimination tree. They need
to be updated by contribution from supernode $\KS$.

The diagonal block $A^{-1}_{\KS,\KS}$ is computed in an independent task denoted by $D_{\KS,\KS}$.

% Operations of each phase were easily broken into sequences of small tasks in order to achieve fine granularity and good flexibility in the scheduling of tasks to multicore or manycore processors. 

In the outer product stage, the update to the lower triangular part of $A^{-1}_{\CS,\CS'}$ may be divided into several updating tasks, and each task corresponds to a submatrix update defined by \eqref{eqn:outerprod}.
%\begin{equation}
%\begin{adjustbox}{scale=0.9}
%$A^{-1}_{\IS,\IS'}\leftarrow A^{-1}_{\IS,\IS'}
%-A^{-1}_{\IS,\KS}\hat{L}_{\KS,\IS'}, \quad \IS\in\CS, \IS'\in\CS', L_{\IS,\IS'}\neq 0,
%$
%\end{adjustbox}
%\end{equation}
%for every $L_{\IS,\IS'}\neq 0, \IS\in\CS, \IS'\in\CS'$  
%below $L_{\KS,\IS'}$ in supernode $\IS'$.
The update to each block $A^{-1}_{\IS,\IS'}$, denoted by $O^{(\KS)}_{\IS,\IS'}$, can be 
computed as an individual task, and all tasks may be executed 
concurrently if there are enough threads (Figure~\ref{fig:outer}).

In the inner product stage, every matrix block  $A^{-1}_{\KS,\IS'}, \IS' \in \CS'$ is updated according to \eqref{eqn:innerprod}.
%\begin{equation}
%\begin{adjustbox}{scale=0.9}
%$
%    A^{-1}_{\KS,\IS'} \leftarrow A^{-1}_{\KS,\IS'} -
%    \sum_{\substack{\IS\in \CS,\\
%    L_{\IS,\IS'\ne 0}}} A^{-1}_{\KS,\IS}
%    \hat{L}_{\IS,\IS'}, \quad \mbox{for}\quad \IS'\in \CS'.
%$
%\end{adjustbox}
%\end{equation}
This corresponds to a block-sparse inner product between $A^{-1}_{\KS,*}$ and $L_{*,\IS'}$.
There are two ways to divide the tasks for the inner product stage. One way is to to treat each matrix product $A^{-1}_{\KS,\IS}\hat{L}_{\IS,\IS'}$ as a separate task, where 
$A^{-1}_{\KS,\IS}\ne 0$ and $\hat{L}_{\IS,\IS'}\ne 0$.
. However, all such tasks will update a common matrix block $A^{-1}_{\KS,\IS'}$, resulting in a ``write conflict'' that must be resolved to maintain thread safety. The conflict can potentially be resolved by using a thread blocking strategy, but this will potentially hinder parallel efficiency.
Our numerical experience indicates that an alternative
solution with a coarser granularity is a more effective strategy.
Since the update to
$A^{-1}_{\KS,\IS'}$ can be regarded as a single task and performed by a single thread, the update of $A^{-1}_{\KS,\IS'}$ for different $\IS'$ can be performed concurrently without conflict. Such tasks are denoted $I^{(\KS)}_{\IS'}$ in Figure~\ref{fig:inner}.

Because $A^{-1}_{\KS,\IS'}$ becomes available long before $A^{-1}_{\KS,\KS}$ 
is computed for $\IS' > \KS$, we decouple
the task that involves using $A^{-1}_{\KS,\KS}$ from other tasks that do 
not depend on the completion of the $A^{-1}_{\KS,\KS}$ block. 
This allows the latter tasks, denoted by $I^{(\KS)}_{\IS'}, \IS' \in \CS'$, to be executed while $A^{-1}_{\KS,\KS}$ is being updated in
task $D_{\KS,\KS}$. 
Tasks that depend on $A^{-1}_{\KS,\KS}$
are denoted by $ID^{(\KS)}_{\IS'}, \IS' \in \CS'$.
%$A^{-1}_{\KS,\KS}$:
%\begin{equation}
%\begin{adjustbox}{scale=0.9}
%$
%    A^{-1}_{\KS,\IS'} \leftarrow A^{-1}_{\KS,\IS'} -
%     A^{-1}_{\KS,\KS} \hat{L}_{\KS,\IS'}, \quad \mbox{for}\quad \IS'\in \CS'.
%$
%\end{adjustbox}
%\end{equation}

%These tasks are denoted $ID^{(\KS)}_{\IS'}, \IS' \in \CS'$.
We rely on task dependency analysis to prevent write conflicts between
task $I^{(\KS)}_{\IS'}$ and task $ID^{(\KS)}_{\IS'}$ by adding a dependency between these two tasks. Therefore, task $ID^{(\KS)}_{\IS'}$ depends on both the completion of tasks $D_{\KS,\KS}$ and $I^{(\KS)}_{\IS'}$.

\begin{figure*}[htbp]
\centering
\begin{adjustbox}{width=.5\linewidth}
\includegraphics{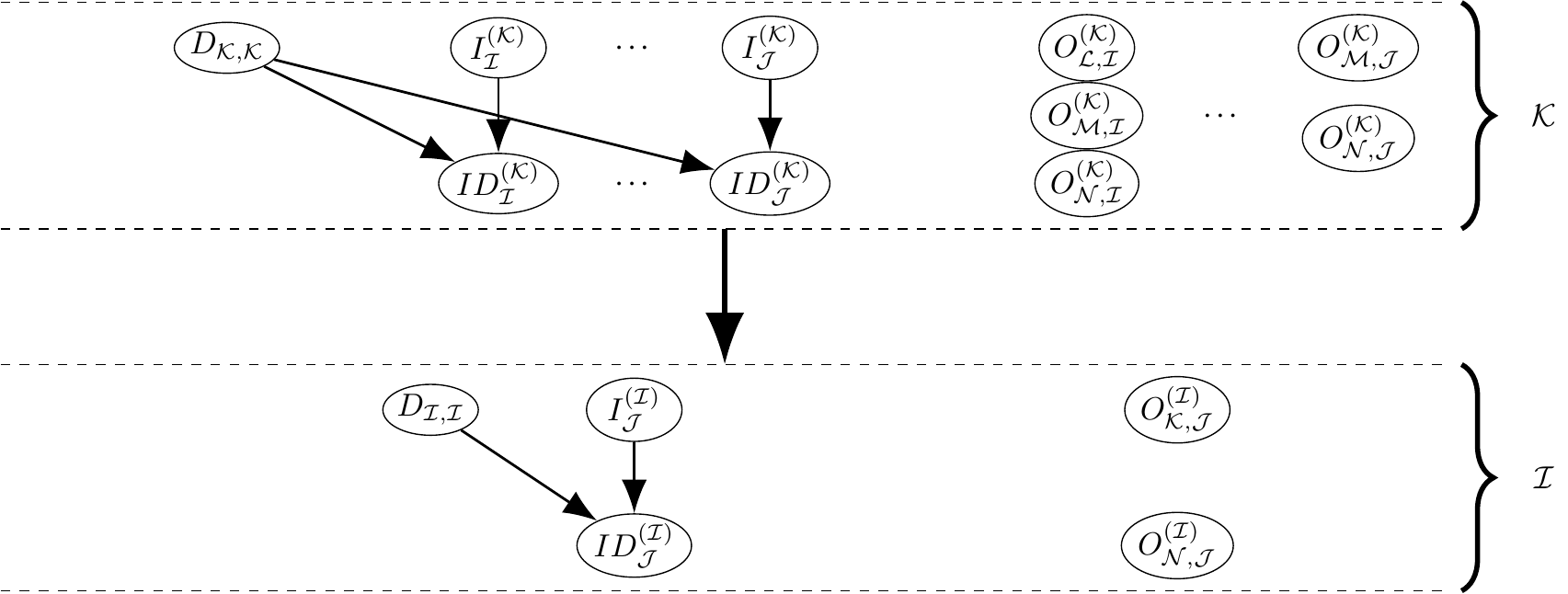}
\end{adjustbox}
\caption{Task dependencies for supernode \KS. \IS is the next supernode that will be processed.\label{fig:deps}}
\end{figure*}

A summary of task dependencies is depicted in Figure~\ref{fig:deps} for two supernodes \KS and \IS.
It should be noted that the outer product stage is completely independent of the inner product as well as the diagonal block update stages.  
 
% Since multiple  so each inner product of block vectors will be finished by a thread, and there is no conflict between threads. 
% , $A^{-1}_{\KS,\IS}\hat{L}_{\IS,\IS'}, \IS\in\KS\cup\CS, \IS'\in\CS'$, is regarded as a task, so update of submatrix $A^{-1}_{\KS,\IS'}$:
 
% will be completed by multiple threads. If the submatrix   $A^{-1}_{\KS,\IS'}$is updated by multi threads at the same time, this type of tasks will cause serious memory access conflict, and it will be not thread safe. Some strategy such as  In another method, Compared to the first method, it is a relatively direct task partition, and the granularity of task is larger. The task number is number of elements in set $\CS'$. In our current implementation, we have adopted the latter, and got a better performance. 

%The update of the diagonal block $A^{-1}_{\KS,\KS}$ also resembles a inner-product procedure between $\hat{U}_{\KS,*}$ and $A^{-1}_{*,\KS}$. Following our treatment of the inner product stage, we treat the update of the diagonal block as one single computational task performed by one thread. 

%It should be noted that the outer product stage is completely independent of the inner product as well as the diagonal block update stages.  
%However, the update from the diagonal block requires the diagonal block $A^{-1}_{\KS,\KS}$ to finish first, hence a task dependency. 

In all three phases above, each task uses the level-3 \textsf{GEMM}  operation  to exploit cache and memory locality. All tasks can then be dynamically scheduled for execution. The sequence of execution is determined by the dependencies among the tasks and the availability of computational resources. In that sense  the execution of the algorithm is performed by asynchronously scheduling the tasks without imposing explicit barriers. This leads to good load balance and parallel scalability. 
%\LL{Maybe not ``fadedown'' in fig. 3? I wondered for a bit whether it was a rendering problem.}
% This asynchronous scheduling results in an out-of-order execution where slow or sequential tasks, such as the update of the diagonal blocks, may be hidden behind parallel. Moreover, asynchronous execution models allow the hiding of latency of access to memory.

\subsection{OpenMP implementation}

% In the following, we discuss how parallelism can be exploited in the left-looking selected inversion algorithm.
% Parallelism between supernodes stemming from the elimination tree of the matrix can be exploited in a similar way than
% in the right-looking selected inversion algorithm currently implemented in PEXSI. This can be done at the distributed
% memory level as well as within a shared memory node.
% When processing each supernode, different levels of concurrency can be expressed. We propose to organize
% computations in three distinct phases: (

In order to describe the tasks and their dependencies, 
we exploit the latest features of OpenMP 4.0, which enables tasks to be described by the \textsf{task} clause, and dependencies described by the additional \textsf{depend} clause.  Simply speaking, the \textsf{depend} clause consists of a list of input and output dependencies for each task, which can be seen as a list of variables or memory addresses from which a given task will read its input data, and to which a task will write its output data.
%\LL{``memory address of an array'' may be confusing to readers not familiar with the matter. Need better explanation or a reference} \MJ{How about now ?}. 
In order to start the execution of a particular task, all dependencies previously submitted to the OpenMP dynamic task queue must have been finished. This restriction allows the program to dynamically set barriers to certain tasks without hindering the execution of the rest of the tasks.  
A pseudo code for describing the task and task dependencies in step 3-7 of Alg.~\ref{alg:selinvluleft} is given in Alg.~\ref{alg:task}. For instance, the task associated with \textsf{depend (out: $\KS)$} means that upon the finish of the task of diagonal block update operations, all tasks that has a dependency clause \textsf{depend (in: $\KS$)} can be executed. In the case when multiple dependency clauses are present, the task can only be executed after all tasks in the  dependency list are complete.

For each supernode $\KS$, we submit all tasks at the beginning. Note that the tasks are just \textit{described} rather than really executed. In particular, the order of which the tasks are submitted do not reflect the order in which the tasks are executed in the OpenMP task scheduling procedure.
%In particular, any memory allocation of temporary arrays etc will be executed on-the-fly only when the actual tasks are executed. 
After all tasks have been submitted, each task will be executed dynamically according to their dependencies. When multiple routes of parallelism are possible, we do not attempt to arrange \textit{a priori} the order in which the tasks are performed. This strategy tends to enhance parallel performance. 
%However, in practice the execution of tasks can happen simultaneously as other tasks are submitted, when the list of tasks is large. 

% The left-looking selected inversion algorithm triggers relatively complex data dependencies. Our goal is to not to use any \textsf{omp barrier} for tasks among the different supernodes. This is because the use of barriers reduces the parallelism among different supernodes in the elimination tree. 

\begin{algorithm}[ht]
\DontPrintSemicolon
\For{$\KS = \mathcal{N}, \mathcal{N}-1, ..., 1$}{
	\#pragma omp parallel\{ \;
	\Indp\#pragma omp single nowait\{ \;
	\Indp\#pragma omp task depend (out: $\KS$) \{\;
	  \hspace{2em} Diagonal block update operations\;
	\}\;

	\For{$\IS' \in \CS'$}{
		\#pragma omp task \{\;
		  \hspace{2em} Outer product operations\;
		\}\;
		\#pragma omp task depend (out: $\IS'$) \{\;
		  \hspace{2em} Inner product operations\;
		\}\;
		\#pragma omp task depend (in: $\KS$, in: $\IS'$) \{\;
		  \hspace{2em} Inner product from Diagonal block operations\;
		\}\;
    }
\Indm\}\;
\Indm\}
}
  \caption{Task based OpenMP implementation of the left-looking selected inversion algorithm.}
  \label{alg:task}
\end{algorithm}

In the future hybrid MPI+OpenMP version of the left looking selected inversion algorithm, each MPI process will be expected to handle multiple supernodes. The use of barrier can hinder the parallel performance in that scenario as well. In order to eliminate the usage of \textsf{omp barrier}, task dependencies must be expressed between communication tasks (to receive data from a remote process for instance) and local computation tasks. If a set of computations is completely local to an MPI process, OpenMP will allow to exploit as much concurrency as possible between these local computations.

\section{Numerical results}
\label{sec:example}

Numerical tests are performed on two platforms from the National Energy 
Research Scientific Computing Center (NERSC). The first one is the Cori supercomputer. Cori computing nodes are each equipped with two 2.3 GHz 16-core Intel
Haswell processors and 128 GB of memory~\cite{cori}. Each core has its own 64 KB 
L1 and 256 KB L2 caches; and there is also a 40 MB shared L3 cache per socket. 

The second platform is an Intel Knight's Corner (KNC) testbed. 
Each computing node is equipped with two Manycore Integrated Core (MIC) architecture card~\cite{babbage}. Each MIC
card has an Intel KNC processor containing 60 cores, with 4 hardware threads per
core and 8 GB memory per card. The 60 MIC cores are interconnected in a high-speed
bidirectional ring. Each MIC core has a 512 KB L2 cache locally with high speed
access to all other L2 caches. All experiments are conducted in ``native'' mode,
meaning that the host processor is not involved in any way in the computations
(as opposed to the ``offload'' mode).

We evaluate the performance of the left-looking selected 
inversion on two sets of matrices. The first group of 
matrices consists of practical electronic structure computation problems
generated from the \siesta~\cite{SolerArtachoGaleEtAl2002} and
\dgdft~\cite{LinLuYingE2012,HuLinYang2015a} software
The second set of matrices is a selection of problems from the widely used University of Florida Matrix Collection~\cite{FloridaMatrix}.
A description of these matrices is given in Table ~\ref{tab.matrices}.

\begin{table*}[htbp]
\centering
\begin{adjustbox}{width=.9\linewidth}
\begin{tabular}{| c | l | c | c | c | }
\hline
\multicolumn{5}{| c |}{Matrices from Electronic structure computations}\\
\hline
Name & Type & $n$ & $nnz(A)$ & $nnz(L+U)$\\
\hline
DGDFT\_ACPNR4\_60 & Phospherene nanoribbon with 1080 atoms from DGDFT& 16000 & 12800000 & 24115200\\
DGDFT\_ACPNR4\_120 & Phospherene nanoribbon with 2160 atoms from DGDFT& 40000 & 40000000 & 76400000\\
DGDFT\_Graphene180 & Graphene with 180 atoms from DGDFT& 3600 & 4480000 & 8040000 \\
DGDFT\_Graphene720 & Graphene with 720 atoms from DGDFT& 14400 & 17640000 & 58480000\\
SIESTA\_MoS2 & MoS2 with 147 atoms from SIESTA& 2401 & 1800995 & 4616651 \\
SIESTA\_DNA & DNA with 715 atoms from SIESTA& 7752 & 2430642 & 8980372 \\
\hline
\multicolumn{5}{| c |}{Matrices from UFL sparse matrix collection}\\
\hline
Name & Type & $n$ & $nnz(A)$ & $nnz(L+U)$\\
\hline
nd3k & ND problem set, matrix nd3k. & 9000 & 3279690 & 30659502\\
nd12k & ND problem set, matrix nd12k. & 36000 & 14220946 & 342223280 \\
raefsky4 & Buckling problem for container model. & 19779 & 1328611 & 13337337\\
ship\_001 & DNV-Ex 2 : Ship structure, predesign model. & 34920 & 4644230 & 31845572\\
smt & 3D model, thermal stress analysis of surface mounted transistor. & 25710 & 3753184 & 29208900\\
\hline
\end{tabular}
\end{adjustbox}
\caption{Characteristics of matrices used in the experiments\label{tab.matrices}}
\end{table*}

The $LU$ factorization is performed by using the \superlu 
software package~\cite{LiDemmel2003}. 
\superlu does not use dynamic pivoting, and as we focus 
first on the symmetric case, our matrices are permuted in a 
symmetric way without taking into account the values of 
matrix entries.

\begin{figure}[htbp]
\begin{adjustbox}{width=\linewidth}
\input{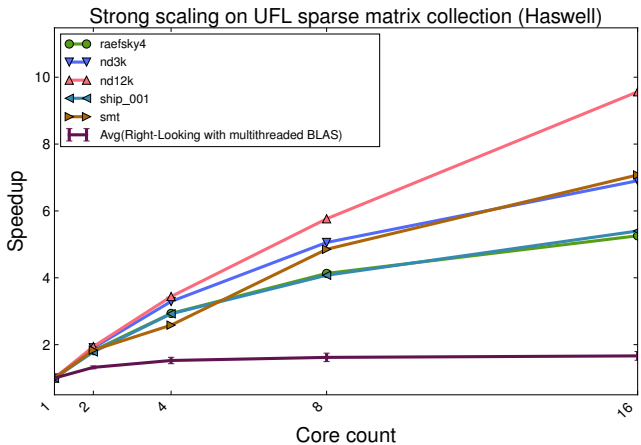}
\end{adjustbox}
\caption{Strong scaling of left-looking selected inversion on Cori (Intel Haswell) for matrices from the University of Florida Sparse Matrix Collection.\label{fig.haswell.ufl}}
\end{figure}

\begin{figure}[htbp]
\begin{adjustbox}{width=\linewidth}
\input{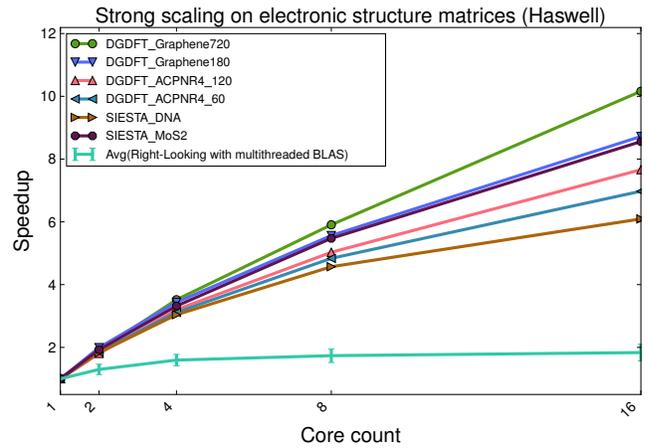}
\end{adjustbox}
\caption{Strong scaling of left-looking selected inversion on Cori (Intel Haswell) for matrices from electronic structure computations.\label{fig.haswell.dft}}
\end{figure}

On the Cori platform, which uses Intel Haswell Xeon 
processors, we observe good strong scalability when
using up to 16 threads. Speedups achieved by the 
left-looking selected inversion algorithm for various
core counts on general sparse matrices are depicted on
Figure~\ref{fig.haswell.ufl}. As a reference, we also
provide the average speedup for the original right-looking
selected inversion algorithm. The associated standard
deviation is represented using error bars. Note that this 
algorithm only leverage parallelism within
BLAS calls. Left-looking selected inversion achieves
speedups ranging from 5.25x to 9.56x using all 16 cores, with an average of 6.84x. 

Speedups achieved on matrices coming from electronic 
structure computations are depicted in 
Figure~\ref{fig.haswell.dft}. Here, speedups range from
6.09x to 10.15x, and an average of 8.02x using all 16 
cores, thus reaching an average parallel efficiency of 
50\%, which is relatively good given the fact that it is a 
sparse matrix computation.

\begin{figure}
\begin{adjustbox}{width=\linewidth}
\input{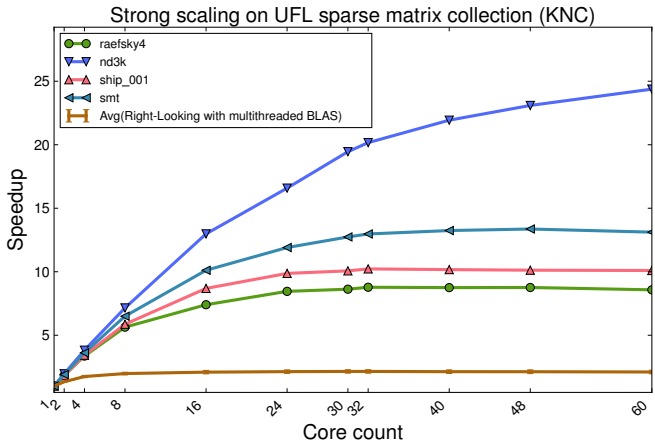}
\end{adjustbox}
\caption{Strong scaling of left-looking selected inversion on Babbage (Intel KNC) for matrices from the University of Florida Sparse Matrix Collection.\label{fig.knc.ufl}}
\end{figure}

\begin{figure}
\begin{adjustbox}{width=\linewidth}
\input{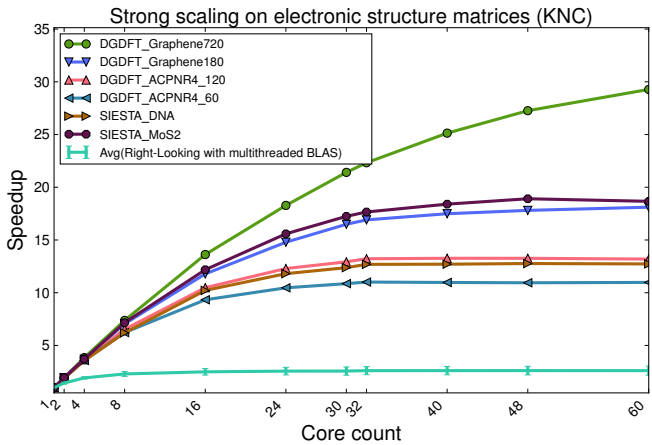}
\end{adjustbox}
\caption{Strong scaling of left-looking selected inversion on Babbage (Intel KNC) for matrices from electronic structure computations.\label{fig.knc.dft}}
\end{figure}

On the Babbage testbed, which uses Intel Knights Corner (KNC) processors, we observe a similar behavior for each
class of matrices. Results are depicted in Figures~\ref{fig.knc.ufl} and~\ref{fig.knc.dft}.
On matrices from the University of Florida collection, 
the speedup results from a 60-core run ranges from 8.58x to 24.38x.
The average speedup of is 14.04x. 
Note that only one thread per core was used in the
experiments.

On matrices generated with \siesta or \dgdft, the speedup results 
from a 60-core run ranges from 10.98x to 29.28x. The average 
speedup is 17.15x. The average parallel efficiency is 28.5\% while 
the maximum efficiency is close to 50\%. 

Altogether, our numerical experiments demonstrate the
practical validity of our approach. Left-looking selected
inversion is able to leverage the parallelism offered
by modern multicore and manycore processors in an efficient 
way. As such, it is a good candidate for a hybrid
MPI + OpenMP implementation that would allow to handle
larger systems.

\section{Conclusion}
We have developed a left-looking variant of selected inversion algorithm,
which can also be viewed as analogous to the right-looking 
factorization algorithm in terms of the  sequence 
of operations. The left-looking selected inversion algorithm 
simplifies task scheduling when multiple tasks are executed simultaneously 
on a parallel machine.  As a first step, we have developed an 
efficient implementation of the left-looking selected inversion algorithm 
for shared memory machines. We demonstrate that, with the task scheduling features provided by OpenMP 4.0, the left-looking selected inversion algorithm can scale well both on the  Intel Haswell multicore architecture and on the Intel Knights Corner (KNC) architecture. The hybrid MPI/OpenMP implementation of the left-looking selected inversion algorithm on multicore and manycore architecture will be our immediate future work.

\section*{Acknowledgment}
This work was partially supported by the National Science Foundation
under Grant No. 1450372 (L. L, Y. Z. and C.Y.), by the 
Scientific Discovery through Advanced Computing (SciDAC)
program funded by U.S. Department of Energy, Office of Science, Advanced
Scientific Computing Research and Basic Energy Sciences (M. J., L. L.
and C. Y.), and the Center for Applied Mathematics for Energy Research
Applications (CAMERA), which is a partnership between Basic Energy
Sciences (BES) and Advanced Scientific Computing Research (ASCR) at the
U.S Department of Energy (L. L. and C. Y.).

\bibliographystyle{IEEEtran}
\bibliography{pselinv}

\end{document}